\documentclass[aps,pre,twocolumn,groupedaddress,superscriptaddress,nofootinbib]{revtex4-1}
\usepackage{graphicx}
\usepackage{amsfonts}
\usepackage{amssymb}
\usepackage{amsmath}
\usepackage{dcolumn}
\usepackage[normalem]{ulem}
\usepackage{enumerate}
\usepackage{bbold}
\usepackage{cancel}
\usepackage{mathtools}
\usepackage{siunitx}
\usepackage{braket}
\usepackage{amsmath}
\usepackage{mdframed}
\usepackage{dsfont}
\usepackage[caption=false]{subfig}

\usepackage{algorithm}
\usepackage[noend]{algpseudocode}

\usepackage{lipsum}

\usepackage[usenames,dvipsnames]{xcolor}
\usepackage[colorlinks,bookmarks=false,citecolor=NavyBlue,linkcolor=OliveGreen,
urlcolor=blue]{hyperref}

\newcommand{\II}{\mathrm{i}}

\newcommand{\up}{\uparrow}
\newcommand{\down}{\downarrow}

\definecolor{gruen}{HTML}{006800}

\def\sw#1{{\color{black}{#1}}}

\begin{document}
\title{Coined quantum walks on the line: Disorder, entanglement and localization}


\author{Louie Hong Yao}
\email{hyao0731@vt.edu}
\affiliation{Department of Physics \& Center for Soft Matter and Biological Physics, MC 0435, 
Robeson Hall, 850 West Campus Drive, Virginia Tech, Blacksburg, Virginia 24061, USA}

\author{Sascha Wald} 
\email{sascha.wald@coventry.ac.uk}
\affiliation{Statistical Physics Group, Centre for Fluid and Complex Systems, Coventry University, Coventry, England}
\affiliation{$\mathbb{L}^4$ Collaboration \& Doctoral College for the Statistical Physics of Complex Systems, Leipzig-Lorraine-Lviv-Coventry, Europe}

\begin{abstract}
\noindent
\sw{Disorder in coined quantum walks generally leads to localization.
We investigate the influence of the localization on the entanglement properties of coined quantum walks.}
Specifically,  we consider quantum walks on the line and explore the effects of quenched disorder in the coin operations.
%
\sw{After confirming that our choice of disorder localizes the walker, we study how the localization affects the properties of the coined quantum walk.}
%
%
%
%
\sw{We find that the mixing properties of the walk are altered non-trivially with mixing being improved at short time scales.}
Special focus is given to the influence of coin disorder on 
\sw{the properties of the quantum state and the coin-walker entanglement.}
\sw{We find that disorder alters the quantum state significantly even when the walker probability distribution is still close to the non-disordered case.
We observe that generically, coin disorder decreases the coin-walker entanglement and that
the localization leaves distinct traces in the entanglement entropy and the entanglement negativity of the coined quantum walk.}

\end{abstract}

\maketitle

\section{Introduction}

Quantum walks have shown significant potential in quantum computing applications 
as a practical tool to build quantum 
algorithms~\cite{She03,Ven12,Kad21,San12,Lecca19,Childs04,Tul08,Mag11,Xia20,Qia21,Aca20,Boe20}.
Even more so, quantum walks are computationally universal
as Childs has shown in his seminal paper~\cite{Childs09}.
Long-standing results of quantum walks outperforming their 
classical counterparts, see, e.g., Refs.~\cite{Tul08,Aha2001,Moo02,Kempe05,Wong18,Far98,Childs04,Ape22},
thus substantiate the promises of quantum technologies to revolutionize 
real-world applications in the near future,
such as high-performance computing or secure communications~\cite{Deu20,Aru19,Gio06,Pez18,Zhong20,Zhang17}.
Despite continuous progress in quantum computing, arguably two of the most influential 
quantum algorithms have been developed decades ago, namely Shor's factorizing algorithm~\cite{Shor} and
Grover's search algorithm~\cite{Grover}. It turns out that designing algorithms with quantum 
advantage is intricate~\cite{Niel10} and thus, it is important to properly understand the fundamental building
blocks of these algorithms which can be formulated \sw{as} quantum walks.

\sw{In this work we will study the influence of of disorder on the entanglement properties of quantum walks.}
Disorder may arise naturally as undesirable error from
faulty quantum operations
but could potentially be used as a tailored resource to improve computational performance
of quantum algorithms.
It is wellestablished that disordered quantum media may lead to wave localization since the
discovery of Anderson localization~\cite{And58}, see also Ref.~\cite{Bik09} for a 
comprehensive overview.
\sw{Localization has been explored} in a variety of setups~\cite{Billy08,Roati08,Kan98,For96,Schw07,Lah08,Wie97,Sto06,Inui04},
even without disorder~\cite{Jahnke08}.
\sw{However, its influence on entanglement witnesses in quantum walks has not been fully explored, with some notable exceptions~\cite{Vieira13,Vieira14,Oma21}. As quantum entanglement is a computational resource~\cite{Chi19}, it is important to understand how entanglement and localization affect one another.
}

\begin{figure}[t]
    \centering
    \includegraphics[width=.8\columnwidth]{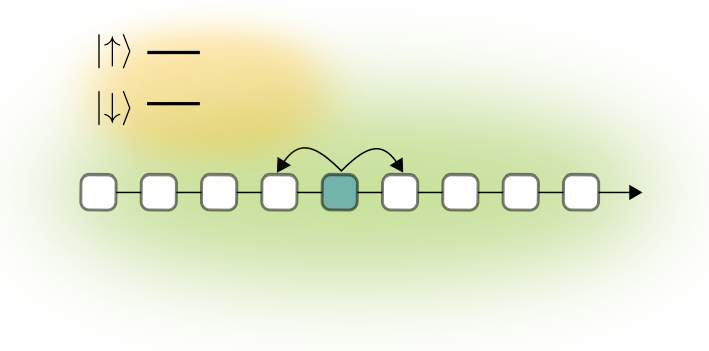}
    \caption{{\bf Illustration coined quantum walk.}
    A CQW is a bipartite quantum system \sw{consisting of} a ``walker''
    and a ``quantum coin.'' \sw{Here,
    the walker takes on discrete positions}
    (green) and the coin is a two-level system (orange).
    The shading illustrates the wave function of each subsystem with the 
    overlap \sw{illustrating} coin-walker entanglement.}
    \label{fig:dtqw}
\end{figure}

To this end, we are interested in coined quantum walks (CQWs) that are routinely described
as \sw{bipartite quantum systems} consisting of a ``walker'' and a ``quantum coin,'' see
Fig.~\ref{fig:dtqw}. 
The walker may occupy sites on a certain geometry (line, circle, general graphs, etc.) and the coin is a 
two level system\footnote{For CQWs on graphs or higher dimensional
geometries, quantum coins with more levels are necessary.} whose
state determines the walker propagation.
Further details will be introduced in Sec.~\ref{sec:model}.
CQWs are generalizations of classical random walks to the quantum realm. Similar to a classical 
random walk on the line, where a coin toss decides whether a walker hops left or right, 
a quantum coin toss decides how the walker spreads.
Here we explore the effects of quenched disorder in the quantum coin toss such that the coin differs on different lattice sites.

In contrast to their classical counterparts, CQWs are generally deterministic since
the underlying \sw{dynamics is unitary}. Further differentiating factors are that
CQWs occupy several sites simultaneously due to quantum superposition and 
the coin-walker interaction yields entanglement.
It has been argued that the natural classical counterparts of CQWs
are \sw{classical random walks} with one-step memory~\cite{Wu19}.
This memory may yield faster spreading than a Markovian classical 
random walk. However,
CQWs still spread asymptotically faster, due to quantum interference~\cite{Wu19}.


Beyond their computational universality, CQWs find applications in, e.g., neural 
networks to capture the structure of graphs~\cite{dernbach2019quantum,Dernbach2019}.
Here, feature-dependent coins are used at different nodes and CQWs can be used to classify graphs.
%
%
%
%
\sw{Hence, coin features may be exploited to perform 
tasks and coin operations play a fundamental role in the character of CQWs which further motivates our study of the effect of quenched coin disorder on entanglement witnesses.}

\sw{
The manuscript is organized as follows. In Sec.~\ref{sec:model}, we introduce CQWs
and the coin disorder we consider.
In Sec.~\ref{sec:proba}, we briefly verify that the coin disorder
yields a walker localization.
We study the influence of the localization on the mixing properties in Sec.~\ref{sec:mix}.
In Sec.~\ref{sec:qs} we explore how the localization affects the quantum state of the bipartite system by considering the state fidelity, the entanglement entropy and the entanglement negativity.
In Sec.~\ref{sec:conclusion}, we summarize our results and discuss some interesting future directions. 
}

\section{The Model}
\label{sec:model}
CQWs are bipartite quantum systems made up from a walker and a quantum coin, see Fig.~\ref{fig:dtqw}. 
Here, we consider arguably the simplest setup where the coin is a two-level system
with corresponding coin Hilbert space
$\mathcal{H}_c =\operatorname{span}\{\ket{\down},\ket{\up}\} = \mathbb{C}^2$
and the walker propagates on the infinite line (alterations are readily introduced).
Thus, the position $x$ of the walker can take on discrete values $x \in \mathbb{Z}$ and 
the walker Hilbert space is 
$\mathcal{H}_w = \operatorname{span} \{\ket{x}\, |\,  x\in\mathbb{Z}\}$. 
The composite system is described by quantum states $\ket{\psi}$ residing in the 
Hilbert space $\mathcal{H} = \mathcal{H}_w \otimes \mathcal{H}_c$,\sw{ i.e., 
$\ket{\psi} = \sum_{x\in\mathbb{Z},\sigma=\up,\down} \psi_{x,\sigma} \ket{x,\sigma}$
where we write $\ket{x,\sigma} = \ket{x}\otimes\ket{\sigma}$}.
This bipartite system undergoes a two-step discrete-time dynamics as follows.
Given a quantum state $\ket{\psi(t)} \in \mathcal{H}$ at time $t$,
first a quantum coin operator
$C$ is applied. $C$ is such that, although it acts on the full Hilbert space, the walker 
occupation probability, i.e., 
\begin{align}
\label{eq:p}
    p_x =\sw{ \left|\braket{x,\up|\psi(t)}\right|^2 + \left|\braket{x,\down|\psi(t)}\right|^2 }
\end{align}
\sw{remains} unchanged. Subsequently, a shift operator $S$ is applied that propagates the walker along the line depending on the state of the two-level system, i.e., 
$\ket{\psi(t+1)} = SC \ket{\psi(t)}$. Time-evolved states are thus deduced from the 
initial state as
\begin{align}
    \ket{\psi(t)} =(SC)^t \ket{\psi(0)}.
    \label{eq:propagation}
\end{align}
Arguably the most studied CQW is the Hadamard walk which is defined by the following 
operators
\begin{align}
    C &= \mathds{1}_w \otimes H    = \mathds{1}_w \otimes \frac{1}{\sqrt{2}}
    \begin{bmatrix}
    1&1\\
    1&-1
    \end{bmatrix},\label{eq:HW_C}
    \\[.25cm]
    S&= 
    \sw{
    \sum_{x\in\mathbb{Z}} \ket{x+1,\up}\bra{x,\up}  + \ket{x-1,\down}\bra{x,\down}. 
    }
    \label{eq:HW_S}
\end{align}
\sw{Here, $H$ is the Hadamard gate.}
For the Hadamard walk, each site is equivalent and a ``fair'' coin toss propagates
the walker. 
The initial coin 
configuration yields the potential for an underlying 
drift in the Hadamard walk since the time evolution is unitary.
\sw{This is circumvented by considering the initial state}
\begin{align}\label{eq:initial}
\ket{\psi(0)} = \frac{1}{\sqrt{2}}\left(\ket{0,\down}+\II\ket{0,\up}\right).
\end{align}
The initial coin state is an eigenstate of the $Y$ gate. Since the Hadamard gate is a 
superposition of the $X$ and the $Z$ gate, this initial coin state ensures that there is no 
drift and the dynamics is symmetric.

\begin{figure}[t]
    \centering
    \includegraphics[width = 0.95\columnwidth]{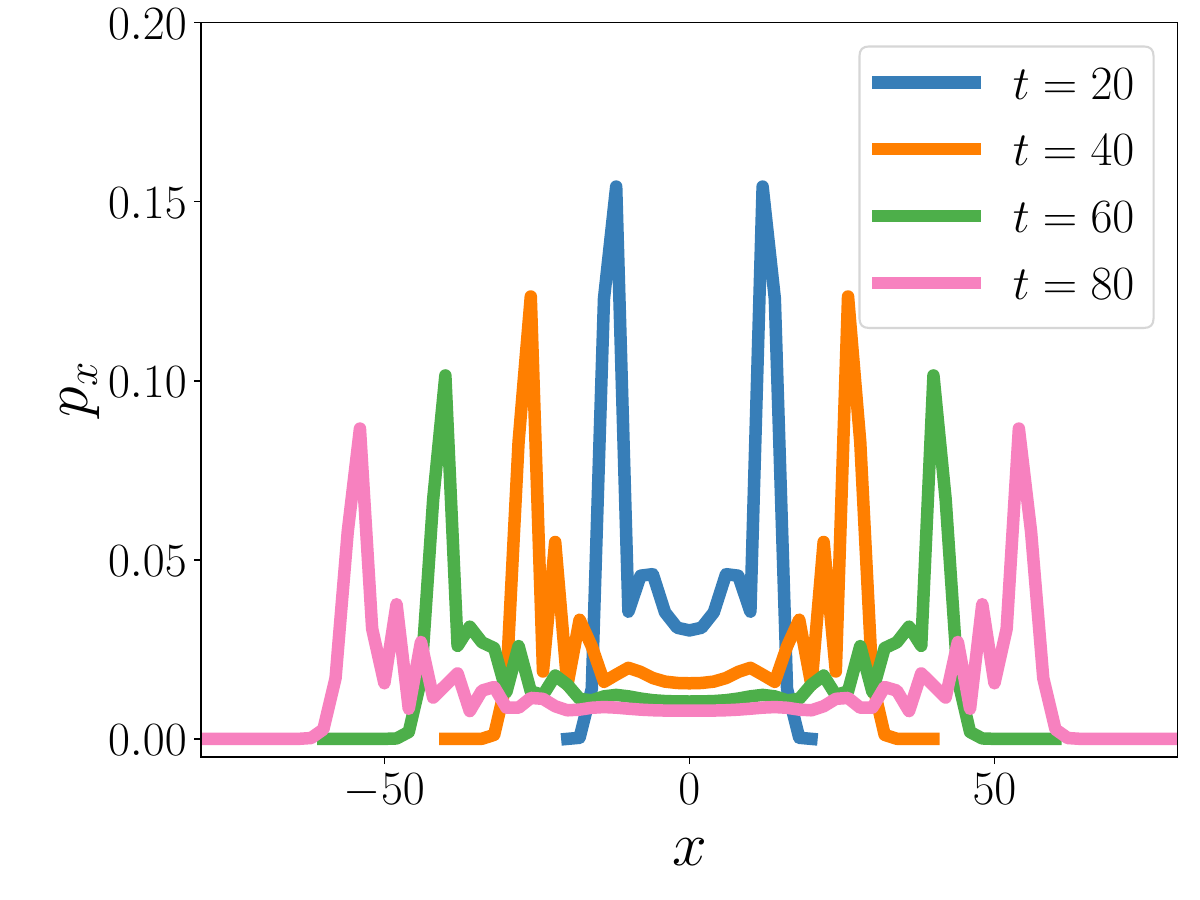}
    \caption{
    {\bf Walker distribution of the Hadamard walk.}
    \sw{We depict the} occupation probability $p_x$ for different
    times $t=20,40,60,80$ for a Hadamard walk with the initial 
    state~(\ref{eq:initial}).
    \sw{We observe that the}
    center probability $p_0$ vanishes 
    and \sw{that the} distribution flattens with time. \sw{The amplitudes of 
    the traveling peaks decay over time}. These results are obtained from the simulation of
    Eq.~(\ref{eq:propagation}).
    }
    \label{fig:HW}
\end{figure}

%
The Hadamard walk and its applications have
been extensively studied, see, e.g., Refs.~\cite{Kon04,nayak2000quantum,Obu15,Tre03,carneiro2005entanglement} to name a few. 
Due to the spatial homogeneity of the coin operator,
the time-evolved wave function can even be analytically derived~\cite{nayak2000quantum}. In Appendix~\ref{app:HW} we recall how to solve the Hadamard walk, following Ref.~\cite{nayak2000quantum}.
For the initial state~(\ref{eq:initial}) we find
\begin{widetext}
    \begin{subequations}
    \begin{align}
    \braket{x,\down|\psi(t)} &= 
    \frac{1+(-1)^{t+x}}{2^{(t+3)/2}} \int_{-\pi}^\pi 
    \frac{dk}{2\pi} e^{-\II kx}
    \left(\sqrt{1+\cos^2 k} - \II \sin k \right)^t\left( 
    1+\frac{\cos k}{\sqrt{1+\cos^2 k}} + \II \frac{e^{-\II k}}{\sqrt{1+\cos^2k}}
    \right),\\[.2cm]
    \braket{x,\up|\psi(t)} &=\frac{1+(-1)^{t+x}}{2^{(t+3)/2}} \int_{-\pi}^\pi 
    \frac{dk}{2\pi} e^{-\II kx}
    \left(\sqrt{1+\cos^2 k} - \II \sin k \right)^t\left( 
    \II \left[1-\frac{\cos k}{\sqrt{1+\cos^2 k}}\right] +  \frac{e^{\II k}}{\sqrt{1+\cos^2k}}
    \right).
    \end{align}
    \label{eq:HW1}
    \end{subequations}
\end{widetext}
The most salient 
feature of the walker occupation probability for the Hadamard walk 
consists of two ballistically propagating peaks emerging
from the origin, see Fig.~\ref{fig:HW}.
The width of these peaks scales asymptotically as $O(t^{1/3})$ and between the peaks the wave function is 
essentially uniformly distributed~\cite{nayak2000quantum}.
Over time, coin and walker become entangled.
This entanglement can be quantified by the entanglement entropy (EE), viz.,
\begin{align}
 S = \operatorname{tr} \rho_c \ln \rho_c.
 \label{eq:EE}
\end{align}
\sw{Here, $\rho_c = \operatorname{tr}_w \rho$ is the reduced density matrix of the coin obtained by 
tracing out the walker.}
For the Hadamard walk, the EE settles to a constant value around 
$0.605$~\cite{carneiro2005entanglement}.\footnote{We considered 
the natural logarithm to estimate this \sw{value which is therefore distinct from values reported elsewhere, e.g., in Ref.~\cite{carneiro2005entanglement}.}}

Certain aspects of disorder in quantum walks have been studied in the past, see, e.g., 
Refs.~\cite{Schr11,Zeng17,Yin08,Rako15,Jack12}.
\sw{Here, we are interested in the effects of disorder in the coin operator
on 
CQWs.}
%
%
%
%
%
%
We consider the gate
\begin{align}
    G(r) = 
     \begin{pmatrix}
     \sqrt{r} & \sqrt{1-r} \\[.2cm]
     \sqrt{1-r} & -\sqrt{r}
     \end{pmatrix}
\end{align}
with $r\in[0,1]$. For certain values of $r$ the gate $G$ reduces to common gates, viz.,
\begin{align}
    G(0) = X, \quad G(1/2) = H, \quad G(1) = Z.
\end{align}
We shall consider CQWs in which the lattice site $x$ has an assigned random number $r_x$
and an associated gate $G(r_x)$. The coin operator is then altered as follows
\begin{align}\label{eq:C}
    C(\{r_x\}) = \sum_{x\in\mathbb{Z}} \ket{x}\bra{x}\otimes G(r_x).
\end{align}
$G(r_x)$ is still composed of an $X$ and a $Z$ gate, viz., 
$G(r_x) = \sqrt{r_x} Z + \sqrt{1-r_x} X$, such that the initial state in Eq.~(\ref{eq:initial})
still yields a symmetric propagation.


Although the coin operator is still in block diagonal form, it does not trivially factorize anymore into a product of two operators acting on $\mathcal{H}_w$ and $\mathcal{H}_c$ separately \sw{due to the site dependent coin operations}.
We introduce a parameter $W$ that controls the disorder strength, i.e.,
\begin{align}
    r_x = \frac{1}{2}\left(1 + W \xi_x\right) .
\end{align}
Here, $\xi_x\in [-1,1]$ are uniformly distributed random numbers. This choice readily
recovers the Hadamard walk for $W = 0$ and a completely disordered walk for $W = 1$.

To analyze the influence of the quenched coin disorder on CQWs, we study 
different disorder realizations.
We write $\{\xi_x^{(i)}\}$, where \sw{$i=1,...,N$} labels the realization.
Each realization $i$ yields a \sw{pure} quantum state $\rho_i(t) = |\psi^{(i)}(t)\rangle\langle\psi^{(i)}(t)|$ \sw{at all times}.
Any system quantity $f$ can be evaluated
\sw{by either first evaluating the density matrix of the ensemble of realizations, viz.,
\begin{equation}
\overline{\rho_i}(t) = \frac{1}{N}\sum_{i=1}^N \ket{\psi^{(i)}(t)}\bra{\psi^{(i)}(t)}
\label{eq:rho}
\end{equation}
and then evaluating the ensemble average $f(\overline{\rho_i})$ or by first evaluating $f$ for a
certain realization and then computing the realization average}
\begin{equation}
\overline{f(\rho_i(t))} = \frac{1}{N}\sum_{i=1}^N f\left(\ket{\psi^{(i)}(t)}\bra{\psi^{(i)}(t)}\right).
\end{equation}
For linear observables, both averages are equivalent, and the expectation values of
an operator $O$ can be defined uniquely as $\braket{O} = \mathrm{Tr}\left(O\overline{\rho_i}\right)$. Hence, the walker occupation probability
$p_x = \operatorname{tr}\left( \ket{x}\bra{x} \overline{\rho_i} \right)$
is directly obtained as
\begin{equation}
p_x(t) = \frac{1}{N}
\sum_{i=1}^N \big|\braket{x,\up|\psi^{(i)}(t)}\big|^2 + \big|\braket{x, \down|\psi^{(i)}(t)}\big|^2.
\label{eq:px}
\end{equation}
\sw{However, for nonlinear $f$, such as the entanglement entropy,
the two averages differ.}

\section{Localization}
\label{sec:proba}
\begin{figure}[t]
    \centering
    \includegraphics[width=.9\columnwidth]{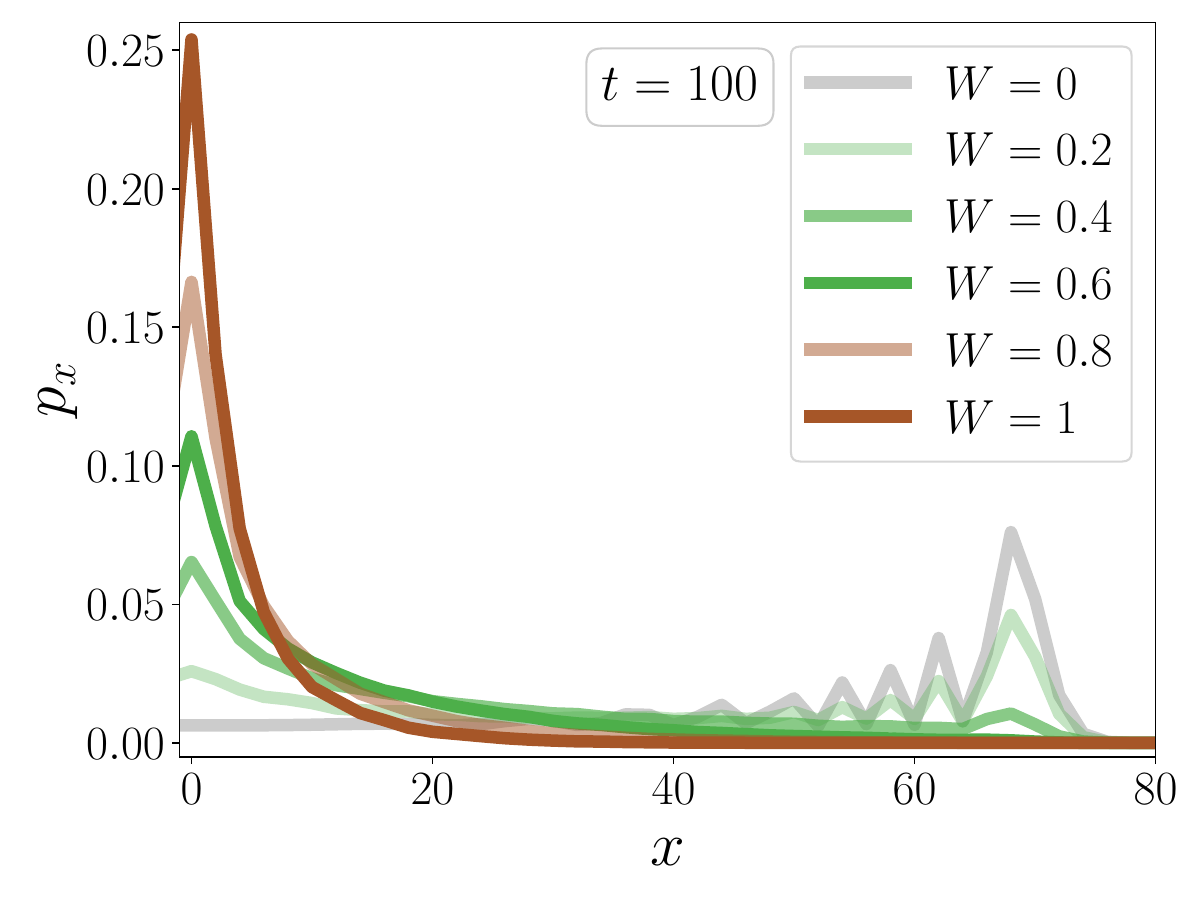}
    \caption{
    {\bf Walker distribution of the disordered CQW.}\sw{ We show the occupation probability
    $p_x$ after $t=100$ time steps}
    for a CQW with \sw{the} disordered coin operator \sw{in} Eq.~(\ref{eq:C}) \sw{and} the initial state~(\ref{eq:initial}) for different 
    disorder strengths $W$.
    \sw{ We observe that the center peak does not vanish in
    the disordered case and increases with the disorder strength.
    Furthermore, the traveling 
    peaks are suppressed with increasing disorder.} These results 
    are obtained by averaging over $1000$ independent coin operator 
    initialization.}
    
    \label{fig:proba}
\end{figure}

\sw{It is expected that the model introduced in Sec.~\ref{sec:model} shows a localization phenomenon. 
In this section we briefly illustrate key indicators of the localization.}

\begin{figure*}[t]
    \centering
    \includegraphics[width=.335\textwidth]{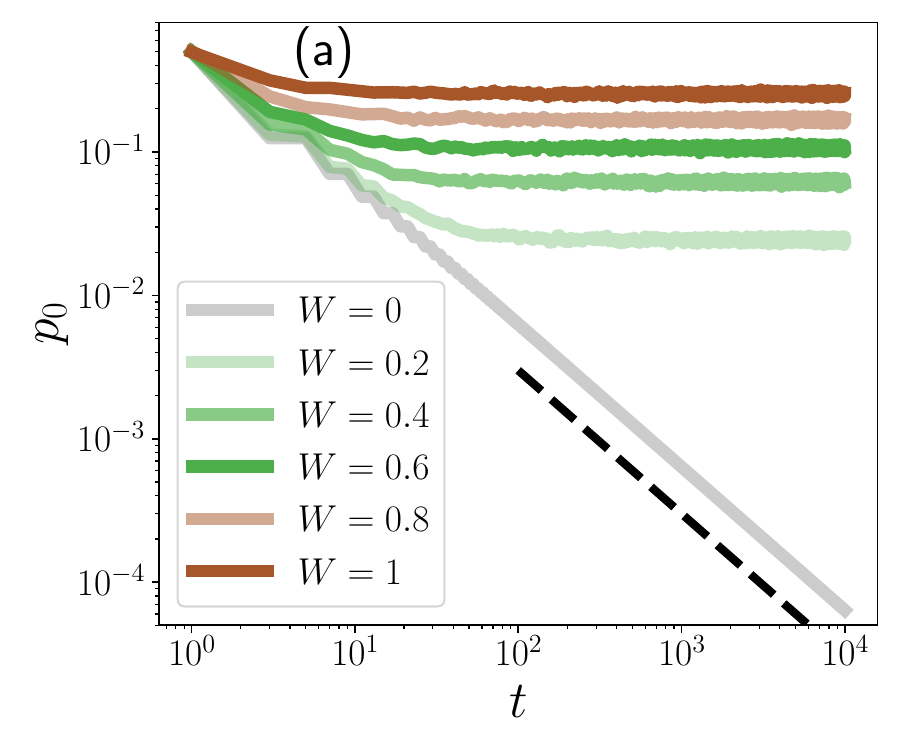}\label{subfig:p0}
    \hspace{-.3cm}
    %
    %
    \includegraphics[width=.335\textwidth]{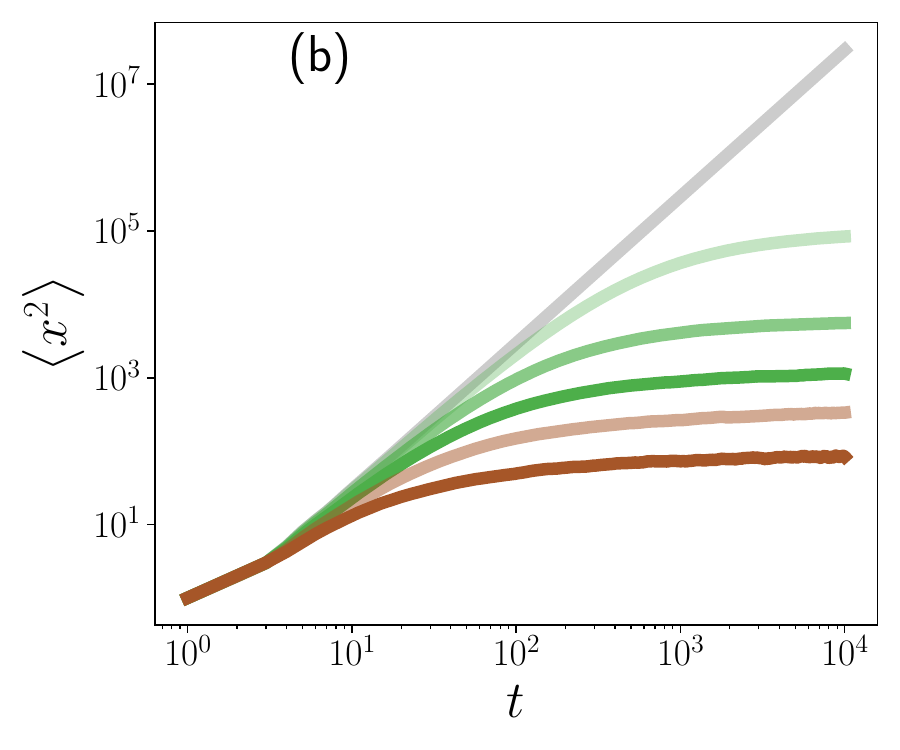}\label{subfig:x2}
    %
    %
    %
    \hspace{-.3cm}
    \includegraphics[width=.335\textwidth]{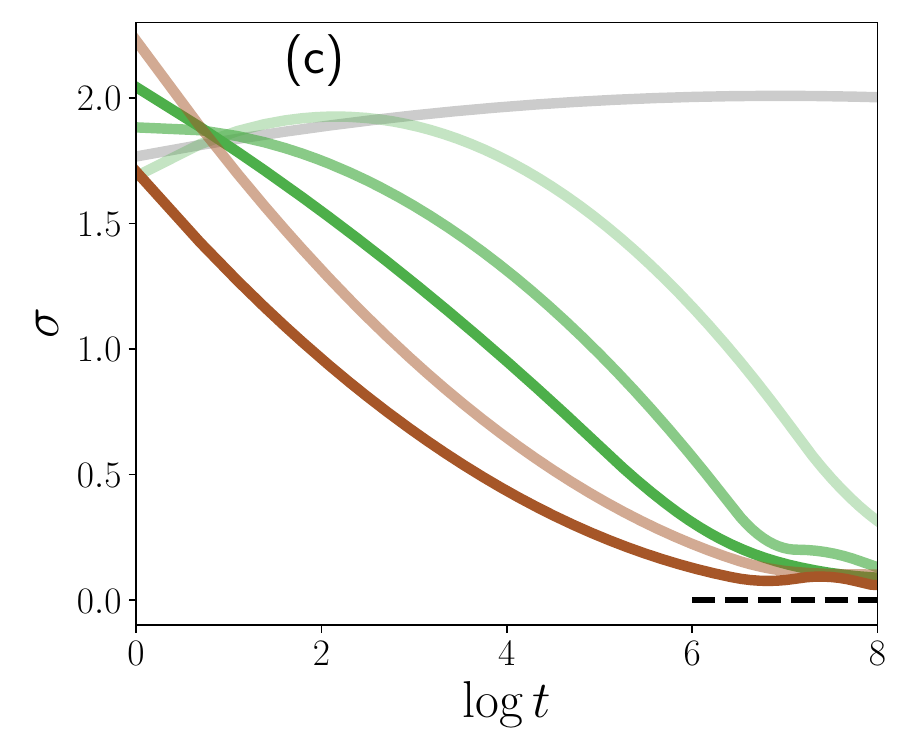}\label{subfig:sigma}
    \caption{
    {\bf Influence of disordered coin operators on walker distribution.} 
    \sw{We consider different properties of the walker distribution and the quantum 
    state for different disorder strengths 
    $W=0$ (gray), 
    $W=0.2$ (lighter green),
    $W=0.4$ (light green),
    $W=0.6$ (green),
    $W=0.8$ (light brown)
    $W=1$ (brown).
    Panel (a) shows the return probability $p_0$. For $W=0$ we 
    observe an algebraic decay as $p_0 \sim t^{-1}$ (as indicated by the 
    black dashed line). Conversely,
    for $W>0$ we see that $p_0>0$ at all times with the plateau 
    value increasing for stronger disorder.
    In panel (b) we show the mean squared displacement. For $W=0$ we find the expected algebraic
    growth. For $W>0$ we observe a significantly slowed down dynamics. 
    In panel (c) this is confirmed by analyzing the growth exponent 
    $\sigma$, see Eq.~(\ref{eq:sigma}). While the Hadamard walk shows $\sigma\to 2$, for $W>0$ we
    consistently find $\sigma\to0$. All results are averaged over $1000$ independent disorder realizations.
    }
    }
    \label{fig:p_disordered}
\end{figure*}

In Fig.~\ref{fig:proba} we \sw{show} the walker occupation probability of a CQW 
[see  Eq.~(\ref{eq:px})] on the infinite line after $100$ time steps for different disorder strengths $W$.
For $W = 0$, i.e., the Hadamard walk, we again observe  the characteristic peaks from
Fig.~\ref{fig:HW}. 
For small to moderate disorder strengths ($W = 0.2, 0.4$) these peaks persist but also a third, central peak 
appears which is qualitatively different from the Hadamard walk. 
For larger disorder strengths, we observe that the 
central peak \sw{is} the dominating feature of the probability distribution indicating that the walker 
is effectively trapped with at most a very slow underlying dynamics.
This behavior is indicative of an Anderson localization in the disordered CQW \sw{and
%
we further study the walker probability distribution in Fig.~\ref{fig:p_disordered}}.
The emergent center peak can be quantified by the return probability $p_0$ of finding
the walker on its initial site $x=0$. 
For the Hadamard walk it is known that $p_0$ decays algebraically as $\sim 1/t$.
In Fig.~\ref{fig:p_disordered}(a), we see this algebraic decay for $W=0$ (note the logarithmic scales).
For different disorder strengths $W>0$, we see that the center peak does not vanish. Instead, we
observe a finite return probability
for $W>0$ and the corresponding plateau value increases with $W$.
Hence, the character of the probability distribution is significantly altered \sw{and the spreading is slowed down}.
In \sw{Fig.~\ref{fig:p_disordered}(b)} we further characterize the slow dynamics \sw{by} considering
the mean squared displacement of the quantum walker, viz.,
\begin{align}
    \langle x^2 \rangle = \sum_{x\in\mathbb{Z}} p_x x^2.
\end{align}
For the Hadamard walk we see that, as expected, $\langle x^2 \rangle$ grows quadratically,
indicating a ballistic dynamics [note again the logarithmic scale in \sw{Fig.~\ref{fig:p_disordered}(b)}].
%
For $W>0$, we \sw{find} that the mean squared displacement deviates from that of the Hadamard walk on a 
finite time scale that decreases with increasing disorder strength and eventually always becomes 
subdiffusive. 
%
\sw{This is evident upon assuming} 
a power-law \sw{behavior}, i.e., $\langle x^2 \rangle \sim t^\sigma$ \sw{which} allows to extract the growth exponent $\sigma$ as
\begin{align}
    \frac{d\log \langle x^2 \rangle}{d \log t} = \sigma.
    \label{eq:sigma}
\end{align}
In \sw{Fig.~\ref{fig:p_disordered}(c)} we extract \sw{the growth} exponent numerically by interpolating the data using 
smoothed cubic splines and evaluating the 
numerical derivative. \sw{For $W=0$, we find asymptotically ballistic spreading, i.e., $\sigma \to 2$ for $t\to\infty$}.
Conversely, a diffusive dynamics would yield $\sigma =1$.
When $\sigma = 0$, 
there is no spreading and instead localization takes place~\cite{Duda22}. For any $W>0$, we see from \sw{Fig.~\ref{fig:p_disordered}(c)} that $\sigma \to 0$.
Hence, 
we \sw{conclude} that the CQW with coin disorder localizes on the length and timescales we explored.
\sw{In Appendix~\ref{sec:bcs} we confirm that our findings also hold true for reflective and periodic boundary conditions.}


\section{Mixing Properties}
\label{sec:mix}

\sw{Mixing in CQWs describes how close the walk is to a certain limiting distribution. This is of fundamental importance in various speedup claims in quantum algorithms~\cite{Cha20}.}
\sw{In this section we study how the localization affects the mixing properties of the DTQW. 
}

\begin{figure}[t]
    \centering
        \includegraphics[width=.9\columnwidth]{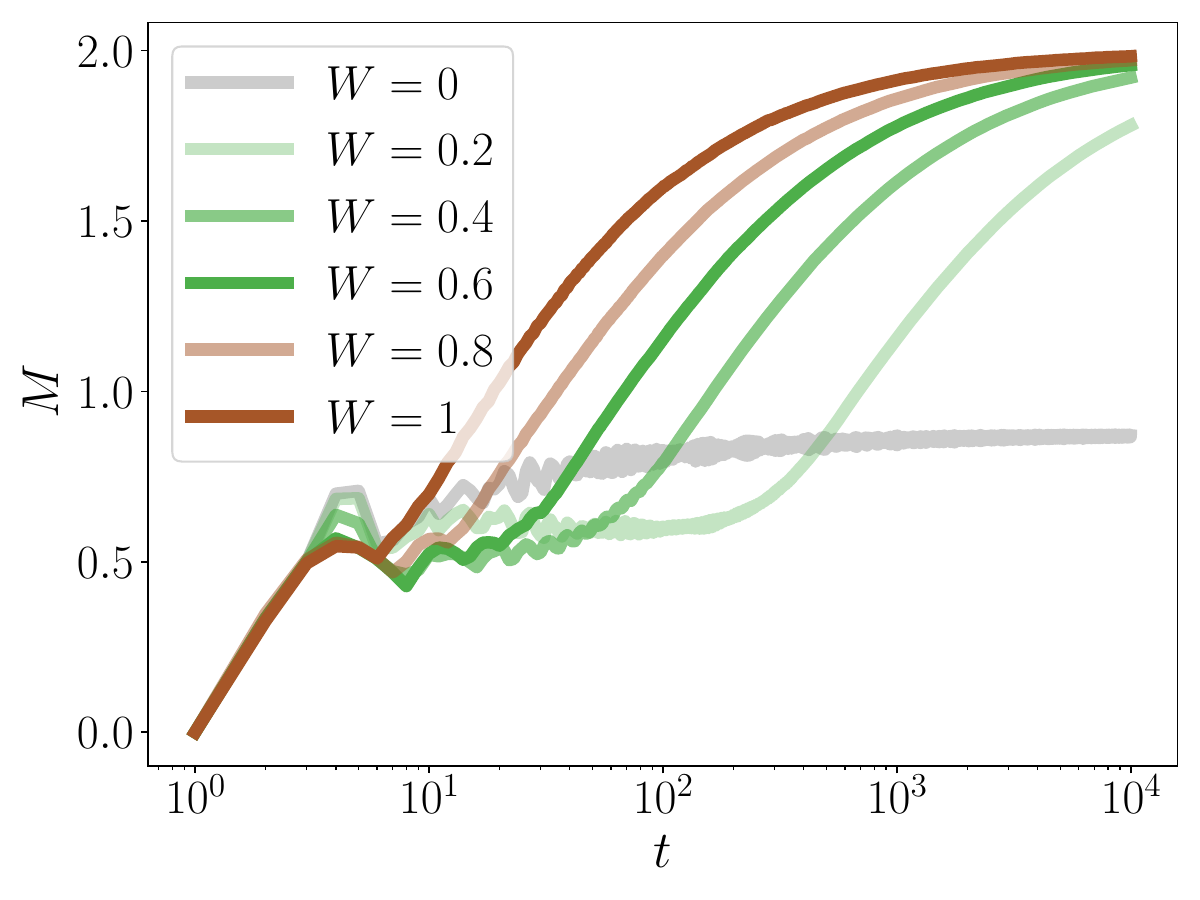}
    \caption{
    \sw{
    {\bf Mixing Ratio.} We show the mixing ration $M$ introduced in Eq.~(\ref{eq:mixing}) for different disorder strengths $W$. For the Hadamard walk ($W=0$) we observe a strong mixing with $M\approx 1$. However, most disordered CQWs with $W>0$ show a stronger mixing in the transient regime than the Hadamard walk. Asymptotically, the disordered CQWs do not have significant overlap with the flat distribution, i.e., $M\to 2$. 
    All results are averaged over $1000$ independent disorder realizations.}
    }
    \label{fig:mixing}
\end{figure}

\sw{To describe mixing in the presence of disorder
we compare the walker occupation probability distributions with the
 flat distribution.}
To ensure a meaningful flat state for the infinite system, we must consider that our initial system 
state implies that at even (odd) times only even (odd) sites can be occupied. 
Hence, we must consider the flat state restricted to the sites that correspond to the parity of the 
current time step and that lay within the \sw{physical light cone $|x|\leq t$} of the dynamics. 
Thus, we refer to 
the following probability distribution
\begin{align}
    p_{\rm flat}(x,t) = 
        \begin{dcases}
        1/(t+1),\ |x|\leq t \ \land \ (t-x) \ \text{even}\\
        0, \quad \text{else}
        \end{dcases}
\end{align}
as the flat distribution at time $t$ for the system on the infinite line. 
The mixing ratio $M$ is then defined by the $1$ norm of the distance
\begin{align}
 M = \left\| p_x - p_{\rm flat}\right\|_1 .
 \label{eq:mixing}
\end{align}
\sw{The mixing ratio satisfies $0\leq M \leq 2$ with $M\approx0$ indicating strong mixing and a probability distribution close to the flat state. Conversely, $M\approx 2$ indicates that the walker probability distribution and the flat state do not have significant overlap.}

\sw{Interestingly, we observe from Fig.~\ref{fig:mixing} that a small amount of disorder $W$ 
initially increases the mixing of the 
quantum walk (e.g., for $W = 0.2$). 
This means that disordered walks are more homogeneously distributed at short time scales than a Hadamard walk.
However, eventually $M\to 2$ for $W>0$ in all cases we considered.}
The fact that \sw{for $W>0$}, all walks tend toward the maximum mixing ratio implies that the \sw{flat state and the walker probability distributions}
do not have significant 
overlap. This means that the walker spreads significantly slower than the physical light cone \sw{$x=t$}.
In turn, the Hadamard walk ($W=0$) keeps up with the light cone and the mixing ratio does not reach 
$M=2$.

\section{Quantum State Properties}
\label{sec:qs}

\sw{Despite that the localization itself is the subject of many studies, the effects on the intrinsic quantum state of the bipartite quantum systems are not yet fully explored.
It is the purpose of this section to analyze these effects of the localization on the quantum states of the CQWs.
First, we compare the quantum state for the disordered system with that of the Hadamard walk.
This will indicate to which extent the disorder affects the quantum state of the CQW.
We then study how the coin-walker entanglement is altered by the presence of disorder.}

\subsection{State Fidelity}
\begin{figure}[t]
    \centering
        \includegraphics[width=.9\columnwidth]{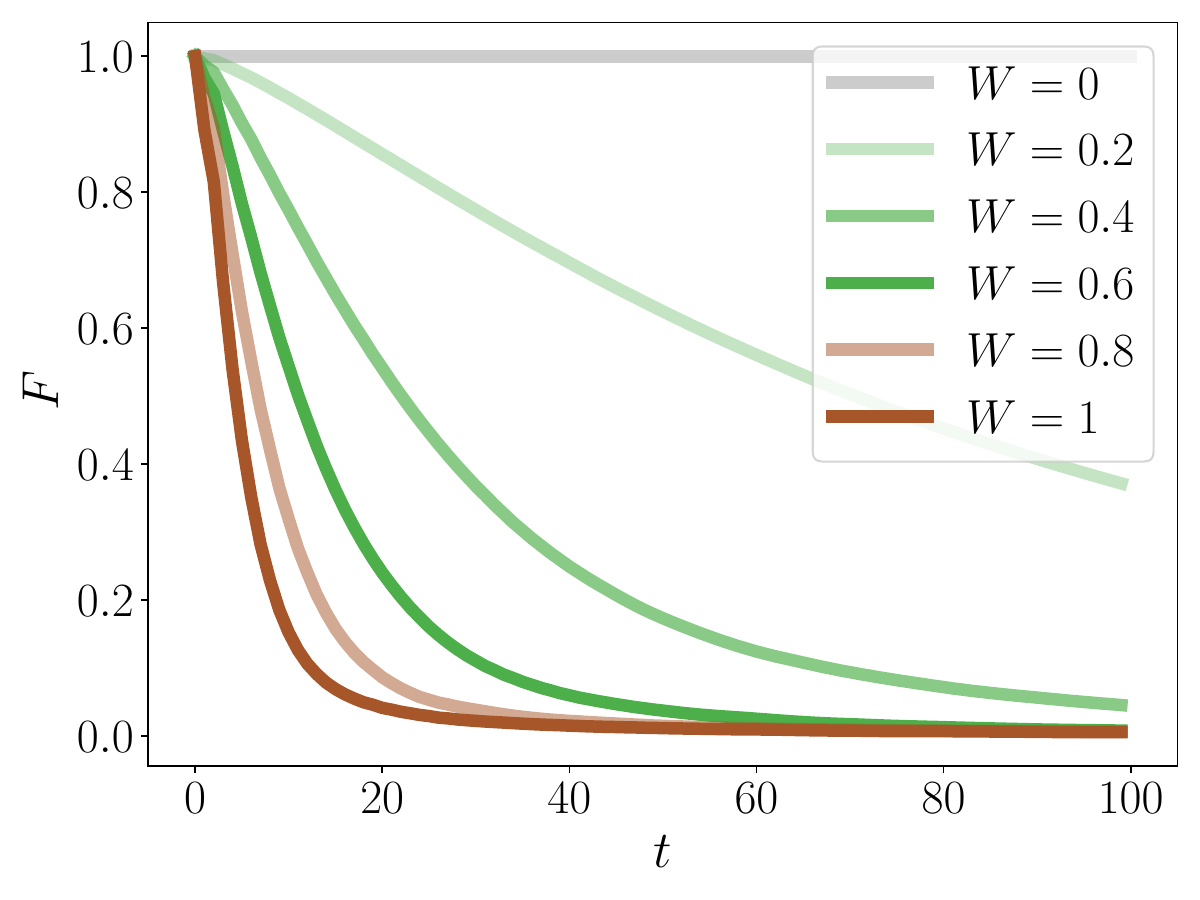}
    \caption{{\bf State fidelity.}
    \sw{
     We compare the quantum state for $W>0$ to that of the Hadamard walk ($W=0$)
     using the
     state fidelity, see Eq.~(\ref{eq:F}). The quantum state quickly deviates from that 
     of the Hadamard walk, even for small $W$.
    All results are averaged over $1000$ independent disorder realizations.
    }
    }
    \label{fig:fidelity}
\end{figure}

\sw{To understand how the quantum state is altered by the coin disorder, we quantify the closeness of the disordered quantum state, see Eq.~(\ref{eq:rho}), to that of the Hadamard walk at the same time, see Eq.~(\ref{eq:HW1}).
For two density matrices $\rho$ and $\sigma$, this can be quantified by the state fidelity $F(\rho,\sigma)=(\operatorname{tr}\sqrt{\sqrt{\rho}\sigma\sqrt{\rho})})^2$.
For a pure state $\sigma = \ket{\psi(t)}\bra{\psi(t)}$, as is the case for the Hadamard walk, $F$ can be written in terms of the individual disorder realizations as follows:
\begin{align}
    F(t) = \frac{1}{N} \sum_{i=1}^N \left| \braket{\psi^{(i)}(t) | \psi(t)} \right|^2.
    \label{eq:F}
\end{align}
$F\approx 1$ indicates that the quantum state $\rho(t)$ has significant overlap with that of the Hadamard walk at the same time. Conversely $F\approx 0$ means that both states do not overlap.
Figure~\ref{fig:fidelity} shows the state fidelity as a function of time for different disorder strengths.
We observe that for all $W>0$ we find $F\to0$, meaning that the quantum state asymptotically has no overlap with the underlying Hadamard walk.
We also see that the state fidelity drops sharply for initial times, implying that even though the walker probability distribution $p_x$ might look similar at short times, the quantum state of the composite system differs strongly from that of the Hadamard walk. This can be observed, e.g., for $W=0.2$ from Fig.~\ref{fig:p_disordered} and Fig.~\ref{fig:fidelity}. 
The larger the disorder strength the faster the state $\rho(t)$ deviates from that of the Hadamard walk.
}

\subsection{Entanglement Entropy}
\label{sec:entanglement}
\begin{figure}[t]
\centering
\includegraphics[width=.95\columnwidth]{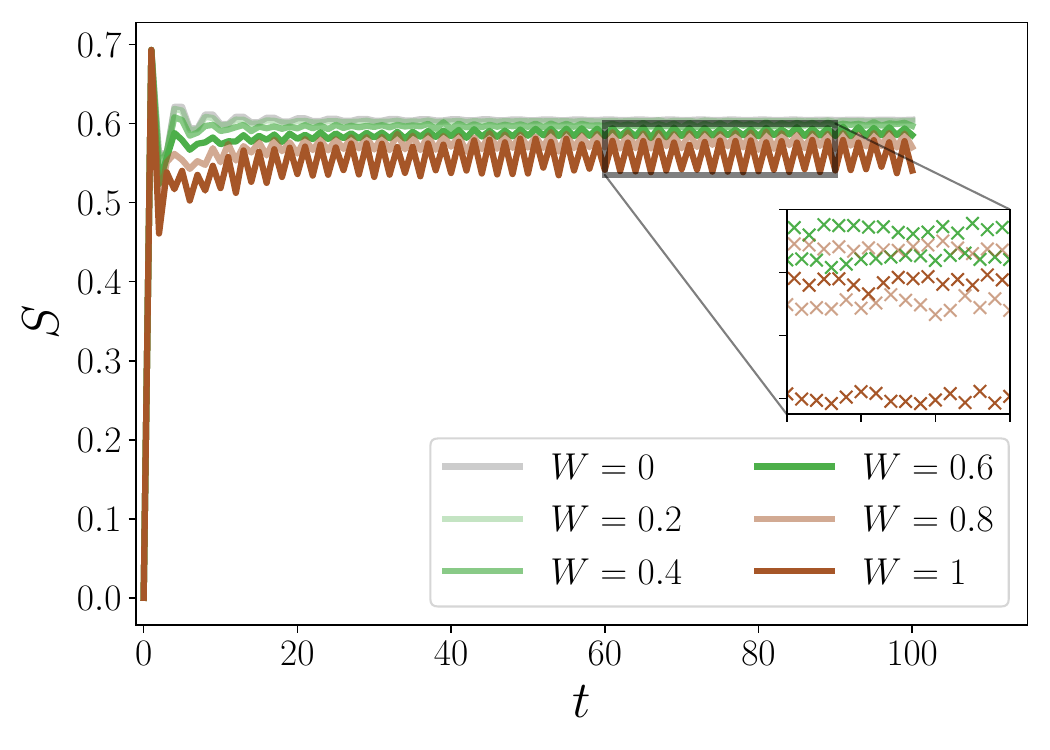}
\caption{{\bf Entanglement entropy.}
For $W>0$, we trace out the walker degrees of freedom for each disorder realization 
and evaluate the resulting EE. The results are averaged over $1000$ independent disorder realizations.
With increasing $W$, oscillations emerge that further support the localization hypothesis. The inset shows
that the EE is rather constant for even (odd) time steps.}
\label{fig:EE}
\end{figure}

\sw{
We have seen that disorder alters the quantum state of the CQW even when localization is not yet apparent (e.g., for $W=0.2$ and $t\simeq 100$).
To further quantify the impact of disorder on the quantum state we study the coin-walker entanglement of the quantum state in the presence of disorder.
}
The initial state given in Eq.~(\ref{eq:initial}) is a pure product state and thus, walker and coin are initially 
not entangled.
Without disorder, the system remains at all times in a pure state and therefore the EE
[see Eq.~(\ref{eq:EE})] is a good entanglement quantifier. As we have explained in Sec.~\ref{sec:model},
in the presence of coin disorder we may consider 
either the ensemble average or the realization average.  The ensemble average describes the statistical 
state of the system but is generally not pure and thus the EE is not a faithful entanglement witness. 
Conversely, every realization stays in a pure state at all times and thus the EE is a good 
entanglement witness for each individual run.
Hence,
to quantify the effects of disorder on the EE we consider the EE for each realization individually 
and average the resulting EEs afterwards.
%
%
In Fig.~\ref{fig:EE} we show the resulting EE.
First, we see that we recover the well-known result for the Hadamard walk for $W=0$~\cite{carneiro2005entanglement}.
Further, we observe that the average EE of the CQW with coin disorder is lower than that of the
Hadamard walk,
indicating that the disorder effectively decreases the entanglement between walker and coin.
With increasing $W$, we also observe that fluctuations are introduced to the entanglement entropy.
Interestingly, these fluctuations do not vanish upon increasing the number of disorder realizations. 
We may understand these oscillations as \sw{a footprint of the} localization phenomenon taking place.
This is because the quantum state cannot be immobile since only those sites can be occupied that
coincide with the time parity (even, odd). Hence, even a state that we consider localized will alter 
between even and odd time steps. The stronger the disorder, the faster the localization occurs such that 
even and odd states \sw{are} picked from the initial transient regime and thus show a significantly 
distinct entanglement character. Hence, the EE for even (odd) steps is rather constant (see
inset in Fig.~\ref{fig:EE}) but the localization
prevents further approaching a unique EE, as is the case for the Hadamard walk.
%
%

\subsection{Entanglement Negativity}

To further investigate the entanglement properties of the CQW with coin disorder, we consider the ensemble 
of realizations. The resulting mixed state $\rho$ is the accurate description of the bipartite quantum system and 
not pure. An entanglement witness for such mixed states is the entanglement negativity which 
is defined as~\cite{Per96}
\begin{align}
    \mathcal{N} = -\frac{1}{2}\left(1-\sum_{i}|\lambda_i'|\right).
\end{align}
Here, $\lambda_i'$ are the eigenvalues of the partial transpose $\rho'$ of 
the density matrix $\rho$, viz.,
\begin{align}
    \rho'_{x\sigma,x'\sigma'} = \rho_{x\sigma',x'\sigma},
\end{align}
with $x,x'\in \mathbb{Z}$ and $\sigma,\sigma' \in \{\up,\down\}$. 
Naturally, if $\rho$ can be written as a sum of 
product states of individual subsystem density matrices,
so can $\rho'$. The normalization $\operatorname{tr}\rho = 1$
is carried over to $\rho'$ such that $\sum_i \lambda_i' = 1$. Thus, in the case that 
$\rho'$ is a sum of product states all $\lambda_i'\geq 0$ such that 
$\sum_i |\lambda_i'| = 1$ and consequently $\mathcal{N} = 0$.
Any negative eigenvalue will indicate a nonseparable quantum state and thus
the presence of entanglement. In this case $\sum_i |\lambda_i'| >1$, and thus $\mathcal{N}>0$.

\begin{figure}[t]
\centering
\includegraphics[width=.95\columnwidth]{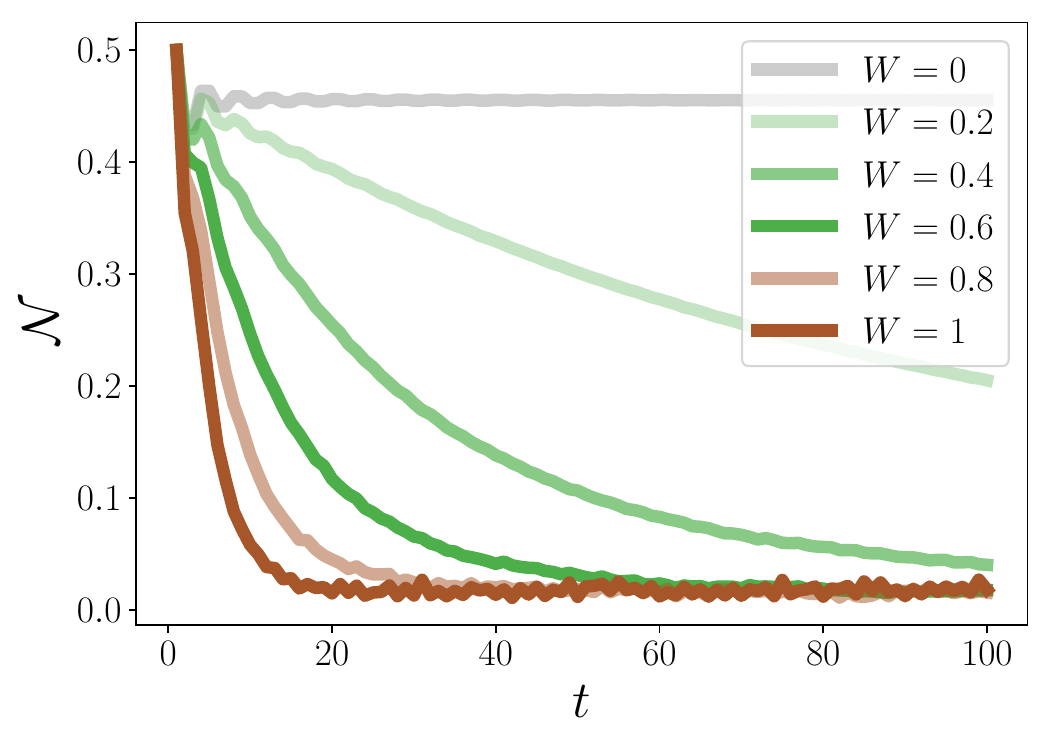}
\caption{{\bf Entanglement negativity.} For different values of the coin disorder $W$, we depict the
entanglement negativity between coin and quantum walker. For $W=0$, we observe $\mathcal{N} \simeq 0.45$ 
for large times clearly indicating an entangled quantum state. Conversely, for $W>0$ we see 
$\mathcal{N} \to 0$. We consider $1000$ independent disorder realizations.}
\label{fig:neg}
\end{figure}

In Fig.~\ref{fig:neg} we show our results for the entanglement negativity for the CQW with coin 
disorder.
For the Hadamard walk, we recover the established behavior of the negativity~\cite{Maloyer_2007}: following 
an initial transient regime, $\mathcal{N}$ settles to a constant value and remains finite. For $W>0$,
we observe \sw{a sharp drop of the entanglement negativity and a continuous downward trend $\mathcal{N}\to 0$.
This behavior hints at an absence of entanglement in the ensemble of initializations.
However, we stress} that this does not mean that the quantum state is not entangled but we may
understand the vanishing of $\mathcal{N}$ as an indication that the quantum state is ``more separable.''
This, again, \sw{is a trace of the localization in the entanglement properties of the walk}: if the state was
separable, then the walker state is unaffected by coin state yielding an effectively frozen walker.

\section{Conclusion and Outlook}
\label{sec:conclusion}
The Hadamard walk, often referred to as quantum analog of the classical random walk, is a well-studied system
that has been considered in a variety of different setups. Here, we addressed the question how 
quenched disorder in the coin operations influences \sw{properties of the quantum state of} CQWs. First, we \sw{confirmed that the disorder we introduce leads to a localization as it is expected.
As indicators for the localization we observed the suppression of the ballistically traveling peaks in the walker occupation probability along with the the emergence of a new, prominent center peak and the sublogarithmic growth of the mean square displacement.}

\sw{Next, we investigated the effects of the localization on the mixing properties of the CQW.
To our surprise, we found an initial regime in which the disordered system is more uniformly distributed than the Hadamard walk. 
However, asymptotically the localization forces the quantum state to be significantly different from the flat distribution.}

\sw{Finally, we explored to the effects of the disorder on the quantum state of the composite system.
The state fidelity revealed that the quantum state with disorder is significantly different from the Hadamard walk, even on time scales on which the walker occupation probability is still similar to the Hadamard walk. 
To further reveal the impact of disorder on the quantum state, we considered the entanglement behavior between the quantum coin and the walker.}
%
%
%
\sw{We} presented two separate approaches the results of 
which point in the same direction. First, we studied the EE. We considered individual 
realizations of the CQW and averaged the EE of each realizations.
We found that the disorder lowers the average EE per run with increasing disorder.
Interestingly we also observed that, with increasing disorder strength 
$W$, the EE shows oscillations that do not decrease
upon increasing the number of disorder realizations. Rather these oscillations are a \sw{witness
of the localization in the CQW} in the following sense: For $W>0$, the walker state is frozen and 
the larger $W$ the faster this happens. But the walker state cannot be equal at all times since the 
system as we have set it up has an underlying even-odd parity. Hence, the system alternates between
two distinct states and these states differ in their entanglement properties yielding increasing 
oscillations for increasing $W$ as the localization happens earlier in the transient regime.

Conversely, to quantify the entanglement properties of the ensemble of realizations, we considered 
the entanglement negativity. The negativity $\mathcal{N}$
is a faithful entanglement quantifier for mixed states, but from $\mathcal{N} = 0$ one cannot automatically
deduce that there is no entanglement present in the system. The negativity for the standard Hadamard walk 
quickly converges to a finite value ${\cal N} > 0$ indicating that the state of the composite quantum 
system at late times is strongly entangled. Upon introducing coin disorder, we observed 
that the negativity quickly decays to zero hinting at a quantum state for the composite system that 
might be separable or for which entanglement is at least not a dominating feature. This would imply that
the quantum coin state does not significantly influence the walker state. Hence, since the coin induces
the walker dynamics, we may interpret this as further \sw{impact} for a walker localization.

\sw{It might first seem counterintuitive that the traces of the localization in the EE and the entanglement negativity of the disordered CQW look qualitatively different as the EE shows oscillations and the negativity decays.
However, both of these quantities test different properties of the quantum state of the composed system.  
The EE measures to extend to which the quantum state of a particular disorder realization is  separable and subsequently we averaged over the realizations.
The negativity on the other hand measures to which extent the full density matrix is separable.
Since both of these measures are non-linear in the quantum state, different aspects of the entanglement properties of the system are tested and therefore the localization shows different imprints on these quantities.}

Despite the apparent simplicity of CQWs, there are a variety of directions with interesting research 
avenues to explore. For example, it would be interesting to further explore the interplay between 
disorder and coin-walker entanglement in the current setup. One might, e.g., consider different initial
conditions that are entangled to varying degrees and see how the disorder affects the entanglement 
over time. Along these lines, one might as well consider a stochastic resetting~\cite{Rose18,Wald21}. This 
could potentially allow to inject entanglement back in the system and might yield a more entangled 
steady state whose localization properties need to be explored carefully.
It would also be interesting to study quenched coin disorder for different topologies. 
Here, one might consider higher dimensional regular lattices or complex networks with higher connectivity.
In these more complicated setups, it might be that localization requires a minimal amount of 
disorder $W>0$. 
In this context, it would be also interesting to explore connections
between disordered CQWs and quantum Hall systems. 
This could be done by linking CQWs with the Chalker-Coddington model~\cite{Chalker88} and 
studying renormalization group approaches to CQWs that have been previously
used in quantum Hall systems, see, e.g., Ref.~\cite{Cain05}.

\section*{Acknowledgments}

We are grateful to L. B\"ottcher, T. Platini and U.C. T\"auber for
valuable discussions that helped advance this project 
and for critical comments on the first 
draft of this manuscript.
We also acknowledge R.A. R\"omer for inspiring discussions
on possible extensions of our work 
\sw{and the referees who made valuable comments on our work.}

\vspace{2cm}

\newpage

\appendix

\section{Solution of the Hadamard walk}
\label{app:HW}
\begin{figure*}[t]
\includegraphics[width=0.33\textwidth]{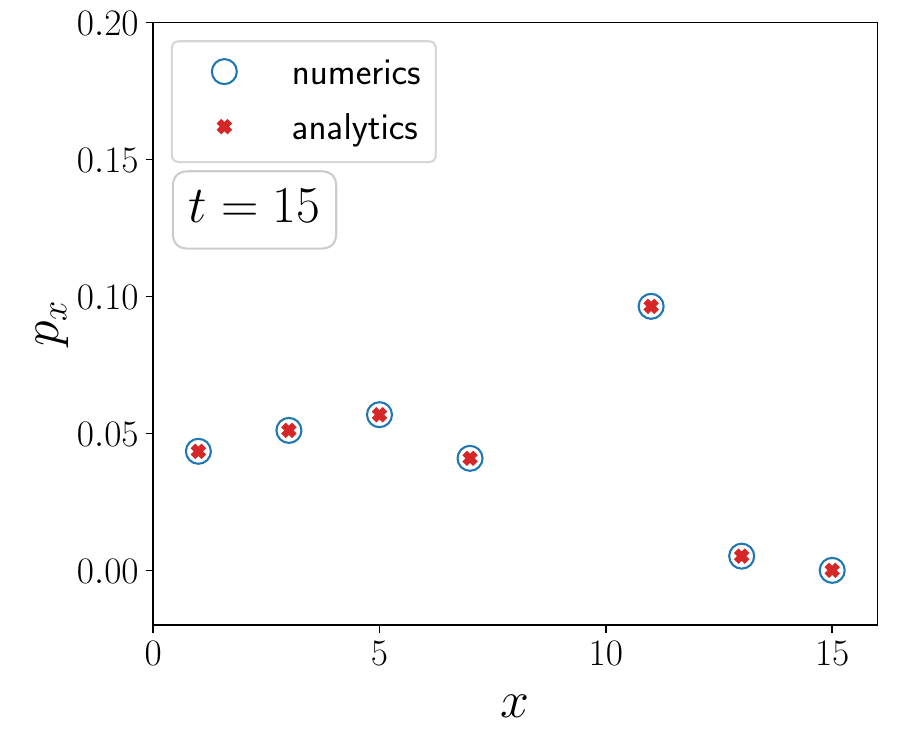}
\hspace{-.3cm}
\includegraphics[width=0.33\textwidth]{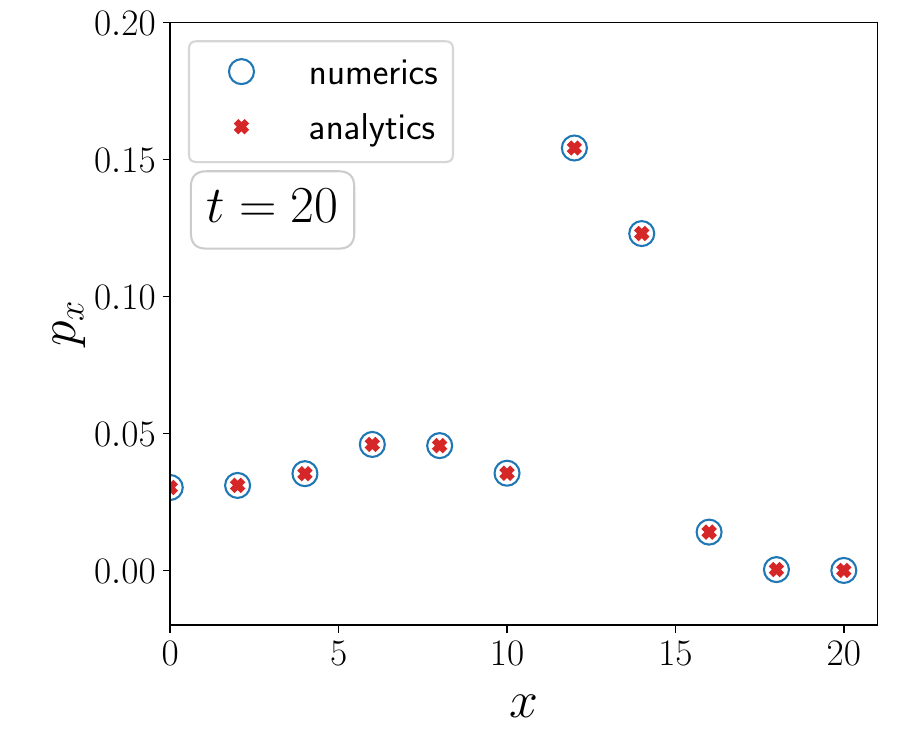}
\hspace{-.3cm}
\includegraphics[width=0.33\textwidth]{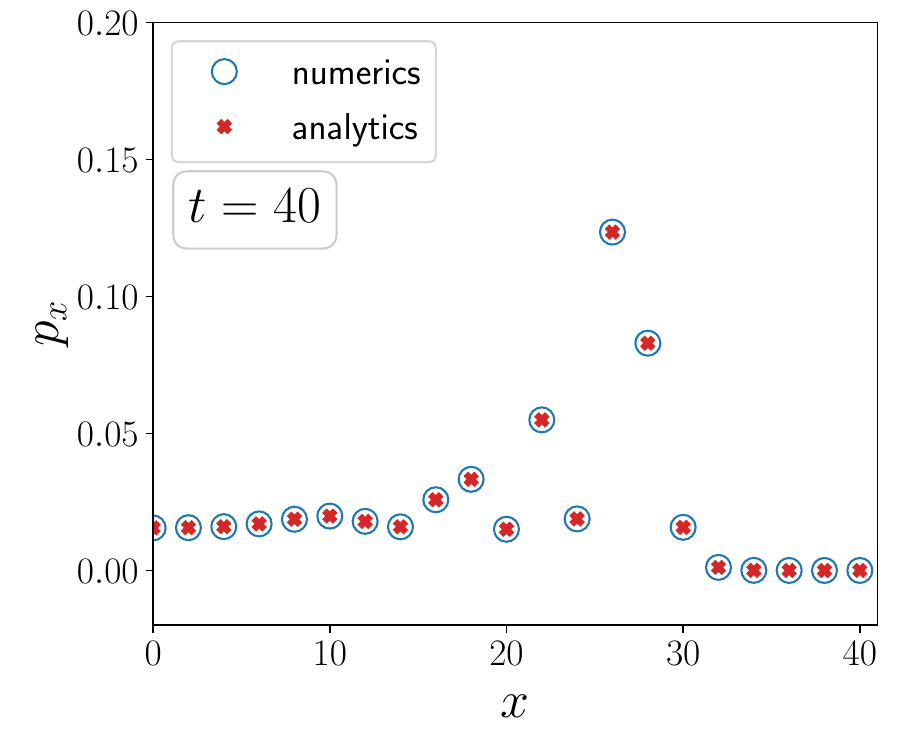}
\caption{{\bf Occupation probability Hadamard walk.} We compare the walker occupation probability $p_x$ 
resulting from the analytical solution of the Hadamard walk [see Eq.~(\ref{eq:HW_ana})] to that
obtained from the numerical evaluation by iteratively applying the coin and shift operator [see Eqs.~(\ref{eq:HW_C}) and~\ref{eq:HW_S})] to the initial state~(\ref{eq:initial}).}
\label{fig:HW_ana_num}
\end{figure*}

Here, we briefly outline the analytical solution of the Hadamard walk, following 
Ref.~\cite{nayak2000quantum}. We write 
\begin{align}
\ket{\psi} &=\sum_x \left(\psi_{x,\down} \ket{\down}+\psi_{x,\up} \ket{\up}\right)\ket{x}
= \sum_{x} \ket{\vec{\psi}_x} \ket{x} 
\end{align}
with the two component vectors $\ket{\vec{\psi}_x} = (\psi_{x,\down},\psi_{x,\up})$.
This allows us to decompose the Hadamard gate into its row components and 
deduce the recursion relation
\begin{align}
    \ket{\vec{\psi}_x(t+1)} = 
    J_+
    \ket{\vec{\psi}_{x-1}(t)}
    +
    J_-
    \ket{\vec{\psi}_{x+1}(t)},
\end{align}
with the matrices
\begin{align}
    J_+ = 
    \begin{pmatrix}
    0&0\\
    \frac{1}{\sqrt{2}}&-\frac{1}{\sqrt{2}}
    \end{pmatrix},
    \quad
    J_- = 
    \begin{pmatrix}
    \frac{1}{\sqrt{2}}&\frac{1}{\sqrt{2}}\\
    0&0
    \end{pmatrix},
\end{align}
such that $H=J_+ + J_-$. Apparently the Hadamard walk is translation invariant, such that we may transform
the recursion relation into Fourier space, viz.,
\begin{align}
 \ket{\Tilde{\vec{\psi}}_k(t)} = 
 \sum_{x} \ket{\vec{\psi}_x(t)} e^{i k x}.
\end{align}
This yields the recursion relation in Fourier space
\begin{align}
    \ket{\Tilde{\vec{\psi}}_k(t+1)}
    &= \mathcal{J}_k \ket{\Tilde{\vec{\psi}}_k(t)}.
\end{align}
Thus, $\mathcal{J}_k\equiv J_+ e^{\II k} +J_- e^{-\II k}$
generates the discrete time evolution of the corresponding Fourier 
component such that $\ket{\Tilde{\vec{\psi}}_k(t)} = \mathcal{J}_k^t \ket{\Tilde{\vec{\psi}}_k(0)}$.
In order to fully determine the dynamics of the Fourier modes it suffices to diagonalize the 
$2\times 2$ matrix $\mathcal{J}_k$. The eigenvalues are readily found, viz.,
\begin{align}
\lambda_{\pm} 
= \frac{1}{\sqrt{2}} \left(\pm \sqrt{1+\cos^2(k)} - \II \sin(k)\right).
\end{align}
Since $\mathcal{J}_k$ is unitary, $\lambda_\pm$ lay on the unit circle in 
the complex plane and we further observe $\lambda_+ = -\lambda_-^*$. Hence 
we may write
$\lambda_+ = e^{-\II \omega_k}$ and $\lambda_- = e^{\II(\pi +\omega_k)}$ with
$\sin\omega_k =\sin(k)/\sqrt{2}$.
%
%
%
The corresponding eigenvectors $\ket{\pm}$ can be found from a straightforward but lengthy calculation, viz.,
\begin{align}
    \ket{\pm} &= 
     \sqrt{ \frac{1}{2}\pm\frac{\cos k}{2\sqrt{1+\cos^2 k}} }
    \begin{pmatrix}
        e^{- \II k}\\
        \pm\sqrt{2}e^{\mp \II\omega} - e^{-\II k}
    \end{pmatrix},
\end{align}
Hence, we may write $\mathcal{J}_k = \sum_{\sigma=\pm} \lambda_\sigma \ket{\sigma}\bra{\sigma}$ and the 
time evolution can be explicitly recovered as 
\begin{align}
    \ket{\vec{\Tilde{\psi}}_k(t)} = \lambda_+^t \ket{+}\braket{+|\vec{\Tilde{\psi}}_k(0)}
    +\lambda_-^t \ket{-}\braket{-|\vec{\Tilde{\psi}}_k(0)}.
\end{align}
For an arbitrary initial coin state $\ket{\Tilde{\vec\Psi}_k (0)} = (a,b)$ we may then write the
time evolved state explicitly in Fourier space as
\begin{widetext}
\begin{subequations}
\begin{align}
    \Tilde{\psi}_k^\down(t) &= 
     \frac{a}{2}\Bigg[
     \left(1+\frac{\cos k}{\sqrt{1+\cos^2k}}\right)e^{-\II\omega_k t}
     +(-1)^t\left(1-\frac{\cos k}{\sqrt{1+\cos^2k}}\right)e^{\II\omega_k t}
     \Bigg]
     +\frac{b}{2} e^{-\II k} \frac{e^{-\II\omega_k t} -(-1)^t e^{\II\omega_k t}}{\sqrt{1+\cos^2k}},
    \\[.5cm]
    \Tilde{\psi}_k^\up(t) &= 
    \frac{a}{2}e^{\II k}\frac{e^{-\II\omega_kt} - (-1)^t e^{\II\omega_k t}}{\sqrt{1+\cos^2 k}}
    +\frac{b}{2} \Bigg[
    \left(1-\frac{\cos k}{\sqrt{1+\cos^2 k}}\right)e^{-\II\omega_k t}
    +(-1)^t\left(1+\frac{\cos k}{\sqrt{1+\cos^2 k}}\right)e^{\II\omega_k t}
    \Bigg].
\end{align}
\end{subequations}
%
%
These Fourier expressions can be readily translated into real space in order to obtain
the walker probability distribution and deduce related quantities such as the mixing ratio. We find


%
%
\begin{subequations}
    \begin{align}
     \psi_{x}^{\down} &= 
     \left(1+(-1)^{t+\sw{x}}\right) \left[
     \frac{a}{2}\int_{-\pi}^\pi \frac{dk}{2\pi}
     \left(1+\frac{\cos k}{\sqrt{1+\cos^2k}}\right) e^{-\II (kx+\omega_k t)}
     +\frac{b}{2}
     \int_{-\pi}^{\pi} \frac{dk}{2\pi}
     \frac{e^{-\II k}}{\sqrt{1+\cos^2k}}e^{-\II (kx+\omega_k t)}
     \right],
     \\
     \psi_{x}^{\up} &= 
          \left(1+(-1)^{t+\sw{x}}\right) \left[
     \frac{a}{2}
      \int_{-\pi}^{\pi} \frac{dk}{2\pi}
     \frac{e^{\II k}}{\sqrt{1+\cos^2k}}e^{-\II(kx+\omega_k t)}
     +\frac{b}{2}
    \int_{-\pi}^\pi \frac{dk}{2\pi}
     \left(1-\frac{\cos k}{\sqrt{1+\cos^2k}}\right) e^{-\II(kx+\omega_k t)}
     \right].
    \end{align}
         \label{eq:HW_ana}
\end{subequations}
\end{widetext}
In Fig.~\ref{fig:HW_ana_num} we compare the occupation probability of the walker obtained from the 
analytical solution to that of the numerical solution for different time steps and we observe 
perfect agreement.

\section{Alternative boundary conditions}
\label{sec:bcs}

For late times, the linear spreading of the light cone makes computational advances
increasingly challenging since the Hilbert space dimension increases rapidly. 
Alternative boundary conditions such as periodic or reflective boundary conditions
naturally limit the Hilbert space dimension and allow more straightforward computational 
approaches. Here we complement our analysis by considering some
properties of the CQW with coin disorder for periodic and reflective boundary conditions.

{\it Periodic boundary conditions}. 
\begin{figure}[t]
    \centering
    \includegraphics[width=\columnwidth]{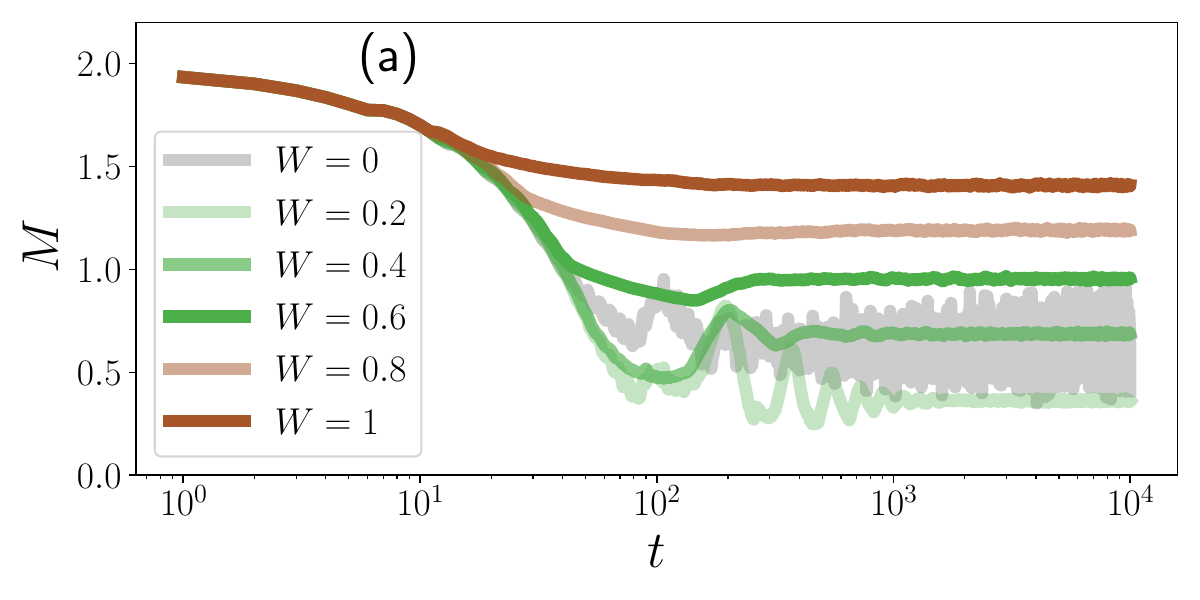}\\
    \includegraphics[width=\columnwidth]{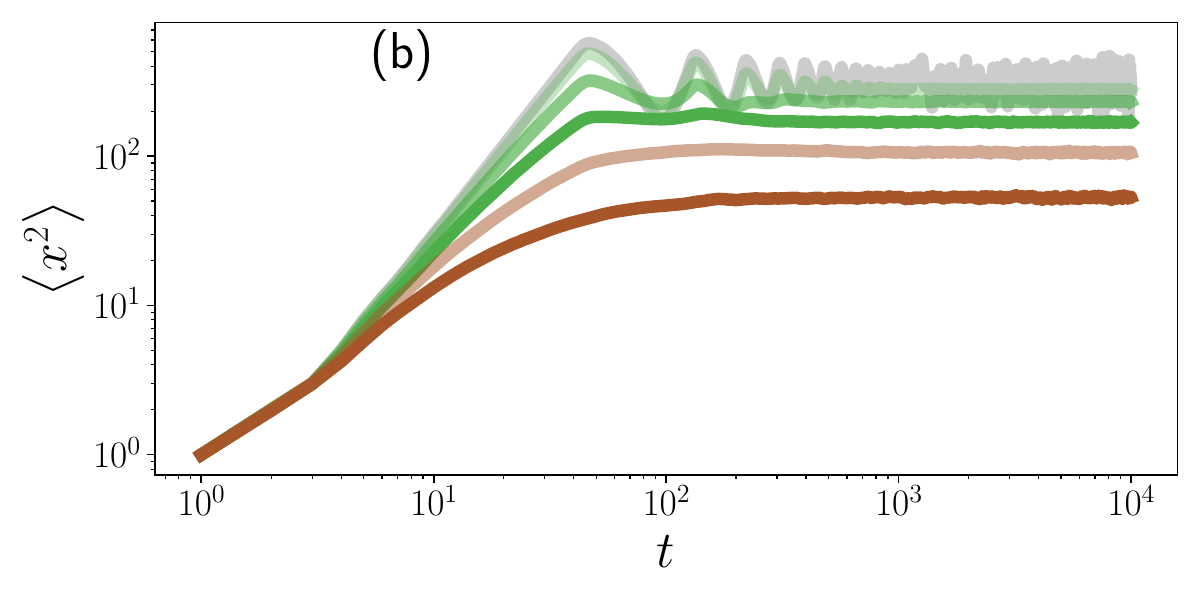}
    \caption{{\bf Periodic boundary conditions.}
    We consider the CQW with quenched coin disorder
    and periodic boundary conditions. To this 
    end we consider a circle with a total of $61$ sites and average over $1000$ independent
    disorder realizations.}
    \label{fig:periodic}
\end{figure}
In Fig.~\ref{fig:periodic} we show the mixing ratio as well as
the mean squared displacement for a system of $61$ sites ($30$ sites in positive and negative direction 
respectively). First, we note that the definition of the flat distribution must be altered to 
evaluate meaningful mixing ratios. This is because the periodic boundary conditions break the even-odd
parity of the CQW due to the boundary hopping which corresponds to an even-even transition. Hence,
we compare the walker probability distribution with $p_{\rm flat}^{\rm (r)} (x) = 1/L$, where $L$ is
the total number of sites. For the Hadamard walk ($W=0$), the initial state [see Eq.~(\ref{eq:initial})] is
far away from $p_{\rm flat}^{\rm (r)}$ and mixing occurs rather fast. Around $t \approx 100 = 
\mathcal{O}(L)$
the mixing ratio peaks since the two ballistic peaks meet and interfere constructively. The spreading
of the peaks with time eventually yields rather constant oscillations and a rather strong mixing.
For small disorder strengths (e.g., $W=0.2$), we see that the CQW mixes better and oscillations are 
notably absent again hinting at a localized state. Upon increasing $W$, we observe that states become less
mixed as we would expect since a single, prominent peak is forming. This is supported by the results 
for the mean squared displacement. For this finite system, there is a natural upper bound and for 
each disorder strength $W>0$, $\langle x^2 \rangle$ saturates at a level below that of the Hadamard walk.

\begin{figure}[t]
    \centering
    \includegraphics[width=\columnwidth]{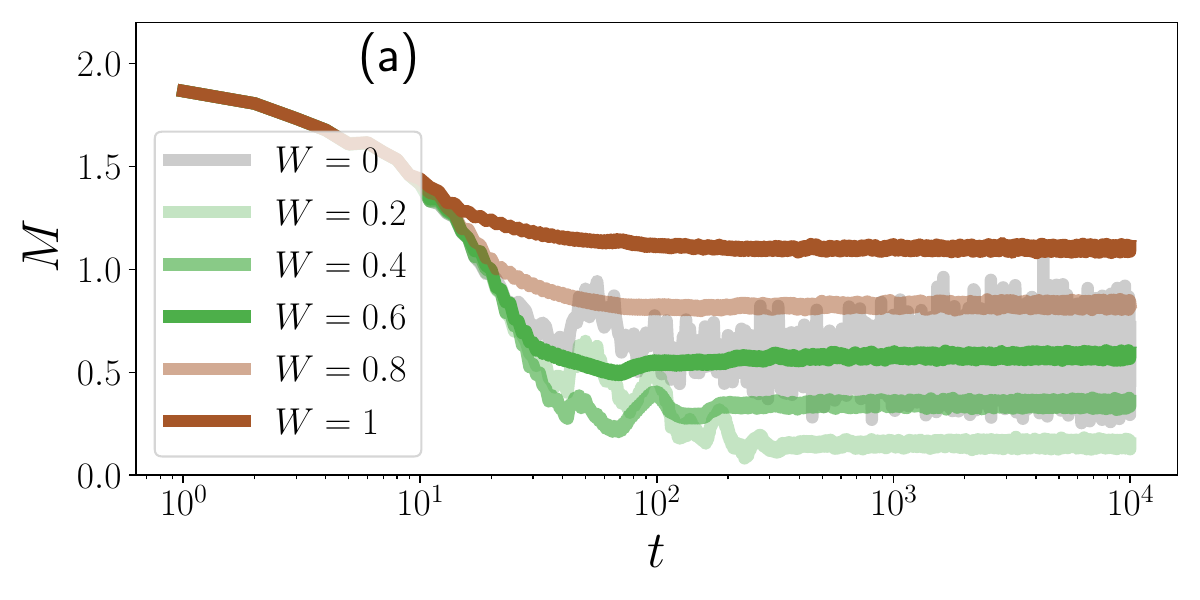}\\
    \includegraphics[width = \columnwidth]{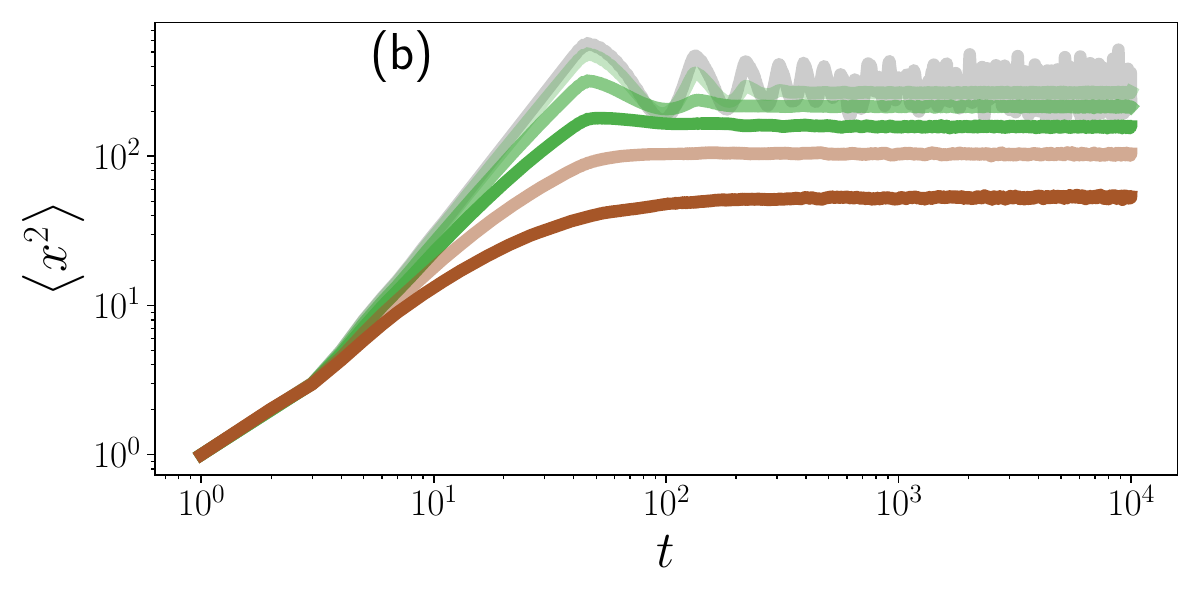}
    \caption{
    {\bf Reflective boundary conditions.}
    We consider the CQW with coin disorder and reflective boundary 
    conditions. To this 
    end we consider a line with a total of $61$ sites and average 
    over $1000$ independent disorder realizations.}
    \label{fig:reflective}
\end{figure}

{\it Reflective boundary conditions}.
In Fig.~\ref{fig:reflective} we show the analogous results to 
Fig.~\ref{fig:periodic} but with reflective boundary conditions. Importantly, reflective boundary conditions again preserve the 
even - odd parity of the CQW but as for the periodic boundary conditions we do not rely on a light cone distribution since the system is finite. 
Qualitatively, the results for reflective boundary conditions 
coincide with those of periodic boundary conditions although the absolute values of mixing and 
mean squared displacement vary slightly.
%


\clearpage


\bibliography{bibliography}

\end{document}